\documentclass{aa}
\usepackage{graphicx, natbib}

\begin{document}
%%%%%%%%%%%%%%%%%%%%%%%%%%%%%%%%%%%%%%%%%%%%%%%%%%%%%%%%%%%%%%%%%%%%%%%%%%%
% Define suitable abbreviations.
\newcommand{\eg}{{\sl e.g. }}
\newcommand{\ie}{{\sl i.e. }}
\newcommand{\etal}{{\rm et al. }}
\newcommand{\NaI}{$\rm Na~\sc I~$}
\newcommand{\CaII}{$\rm Ca~\sc II~$}
\newcommand{\heii}{He{\sc ii}}
\newcommand{\civ}{C{\sc iv}}
\newcommand{\siiv}{Si{\sc iv}}
\newcommand{\chis}{$\chi^{2}~$}
\newcommand{\chir}{${\chi}_{\nu}^{2}$}
\newcommand{\kms}{$\,$km$\,$s$^{-1}$}
% Define some commonly used notations, subscripts and superscripts.
\newcommand{\rhounit}{\mbox{$M_\odot \,$pc$^{-3}$}}
\newcommand{\vunit}{\mbox{km\,s$^{-1}$}}
\newcommand{\Me}{\mbox{$M_\oplus$}}
\newcommand{\Msun}{\mbox{M$_{\odot}$}}
\newcommand{\Rsun}{\mbox{R$_{\odot}$}}
\newcommand{\Lsun}{\mbox{L$_{\odot}$}}
\newcommand{\Mdot}{\mbox{M$_{\odot}$~yr$^{-1}$}}
% Define commands for `less than or approximately equal to' \ltsimeq
%                     `greater than or approximately equal to' \gtsimeq
\newcommand{\ltsimeq}{\raisebox{-0.6ex}{$\,\stackrel 
        {\raisebox{-.2ex}{$\textstyle <$}}{\sim}\,$}} 
\newcommand{\gtsimeq}{\raisebox{-0.6ex}{$\,\stackrel
        {\raisebox{-.2ex}{$\textstyle >$}}{\sim}\,$}} 
\newcommand{\prpsimeq}{\raisebox{-0.6ex}{$\,\stackrel
        {\raisebox{-.2ex}{$\textstyle \propto $}}{\sim}\,$}}

% Define some useful abbreviations for magnitude systems
\newcommand{\UBV}{$U\!BV$}
 \newcommand{\JHK}{$J\!H\!K$}
\newcommand{\JHKL}{$J\!H\!K\!L$}
\newcommand{\UBVRI}{$U\!BV\!RI$}
\newcommand{\AV}{\mbox{$A_V$}}

\defcitealias{Ioannou01}{Paper~I}

%%%%%%%%%%%%%%%%%%%%%%%%%%%%%%%%%%%%%%%%%%%%%%%%%%%%%%%%%%%%%%%%%%%%%%%%%%%

\title{Understanding the LMXB X2127+119 in M15}

\subtitle{II. The UV data}

\author{Zach Ioannou,
       \inst{1,2}
        L. van Zyl,
       \inst{3}
        T. Naylor,
       \inst{1,4}
        P.A. Charles,
       \inst{5}
        Bruce Margon,
       \inst{6}
        Lydie Koch-Miramond,
       \inst{7}\\
        S. Ilovaisky
       \inst{8}
        }

\offprints{zac@astro.as.utexas.edu}

\institute{$^{1}$Department of Physics, Keele University, Keele, Staffordshire, ST5 5BG, UK\\
           $^{2}$Department of Astronomy, University of Texas at Austin, C-1400, Austin, TX 78712, USA\\
           $^{3}$Department of Astrophysics, Oxford University, Nuclear Physics Lab., Keeble Road, Oxford, OX1 3RH, UK\\
           $^{4}$School of Physics, University of Exeter, Stocker Road, Exeter, EX4 4QL, UK\\
           $^{5}$Department of Physics and Astronomy, University of Southampton, Southampton, SO17 1BJ, UK\\
           $^{6}$Space Telescope Science Institute, 3700 San Martin Drive, Baltimore, MD 21218, USA\\
           $^{7}$Service d'Astrophysique, DSM/DAPNIA/SAp, CE Saclay, 91191 Gif-sur-Yvette Cedex, France\\
           $^{8}$Observatoire de Haute-Provence, 04870, Saint-Michel-l'Observatoire, France
           }

\date{Received}

\abstract{ 
We present {\it HST} UV  observations of the high-inclination low mass
X-ray  binary AC211  (X2127+119),  which is  located  in the  globular
cluster M15 (NGC 7078).  We have discovered a {\civ} P
Cygni  profile in  this system,  which  confirms the  existence of  an
outflow from AC211. The outflow  velocity as measured from the P Cygni
profile is  ${\simeq}1500$~km s$^{-1}$. We calculate that  the mass lost
through this wind is too small to support a large period derivative as
favoured  by  \cite{Homer98}.   Using  new  X-ray
observations we  have revised the ephemeris  for AC211 and  we find no
evidence in support  of a period derivative. The  UV spectrum exhibits
several  absorption features  due to  O, Si  and C.   The  very strong
{\heii}  line at  1640{\AA}  is  not seen  to  modulate strongly  with
orbital phase,  suggesting its origin lies  in the outer  parts of the
system.   
In  contrast,  the eclipse of the  UV  continuum is
short compared with the X-ray  and optical  eclipses.

}

\authorrunning{Zach Ioannou et al.}
\titlerunning{Understanding the LMXB X2127+119 in M15}

\maketitle

%------------------------------------------------------------------------------------------

\section{Introduction}

The star AC211 was first suggested as the optical counterpart of
the X-ray source X2127+119 by \cite{Auriere84}, an association
primarily based on its ultraviolet excess ($U-B$=-1.4).
Their results were spectroscopically confirmed by \cite{Charles86}, making  
this the first optical counterpart of a globular cluster 
X-ray source to be identified.

Since then, there have been several studies of AC211 using 
optical spectroscopy \citep{Naylor88, Bailyn89, 
Ilovaisky89} as well as UV observations 
with the {\it International Ultraviolet Explorer (IUE)} \citep{Naylor92}.
All these suffered from heavy contamination due to background 
cluster stars. More recently \cite{Downes96}
acquired observations with the {\it Faint Object Spectrograph (FOS)}
on board the {\it Hubble Space Telescope (HST)} and were able 
to resolve AC211 from the background stars in the cluster. 
They confirmed that several spectral features observed 
with {\it IUE} are intrinsic to AC211. However, their spectra were 
of relatively low resolution.
In addition, \cite{Callanan99} have recently 
detected EUV emission from M15 which could originate from AC211. 
If confirmed, this system would be the first persistent LMXB source 
detected at EUV wavelengths.

It has been recognized from the very earliest optical observations
\citep{Auriere84} that AC211 may be an eclipsing
system, and this model has been slowly elaborated, and the system is
now generally accepted to be an Accretion Disc Corona (ADC) source,
in which the compact object is hidden behind the rim of the accretion
disc, and the majority of the X-ray come from a scattering corona
\citep[{\it e.g.\ \rm}][]{wh82}.
The largest problem in this interpretation was the observation of two
type I X-ray bursts \citep{Dotani90, s01}, 
which implied a direct line of sight to the surface 
of the neutron star.
This puzzle was recently solved by \cite{White01} who
obtained {\it Chandra} images of the core of M15, which revealed the
presence of a second X-ray source 2.7'' from AC211.
This source (M15 X-2) was completely unresolved by the X-ray instruments flown prior to 
{\it Chandra}, and is now presumed to have been the source of the
X-ray bursts.

In this paper we present two sets of {\it HST} observations with the
{\it Space Telescope Imaging Spectrograph (STIS)} instrument. 
We show that the strong {\heii} emission must originate in the ADC, 
while most of the continuum emission originates in a region very close 
to the centre of the disc. In addition we present the first detection
of a {\civ} line in this system, which exhibits a P Cygni profile and
thus confirms the existence of a wind.

%------------------------------------------------------------------------------------------

\section{Observations \& Data reduction}

Observations of AC211 were taken with the
FUV-MAMA detector of the {\it HST} STIS spectrograph during two ``visits''.
Although the spacecraft is pointed towards the target for the duration
of each visit, the observations are broken up by Earth occultations.
Visit 1 on 1998 July 7 comprises 15 spectra with 
equal exposure lengths of 600 seconds. Visit 2 occurred 11 days later
on 1998 July 18 and again we obtained 15 spectra of 600 second 
exposures. The grating used in both visits was the G140L, with a slit 
52'' long and 0.5'' wide. However, only ${\simeq}25$'' of the long slit 
projects onto the detector. This configuration produced spectra in the 
range of 1150-1730{\AA} with a dispersion of 1.2{\AA} per pixel at 
1440{\AA}. There were also a number of spectra from other sources 
recorded on the MAMA detector. The alignment of the slit on the sky
was different for each visit and therefore apart from AC211 the 
additional spectra on the MAMA detector correspond to different stars 
for each visit.

The data returned from STScI included the fully reduced one
dimensional spectra, as well as the two dimensional reduced spectral
images. However, we noticed that in some cases the reduced data showed 
a poor subtraction of the geocoronal Ly${\alpha}$ emission and in a 
single case the reduced spectrum corresponded to a different object. 
We thus re-extracted the data. We took the flat fielded
and bias corrected data from the STScI pipeline and used an optimal
extraction technique \citep[{\it e.g.\ \rm}][]{Horne86} to extract the spectra from the
2-dimensional images.

We used the wavelength calibration determined by STScI. We then
created a response curve by dividing one of our extracted spectra with 
a reduced spectrum from the STScI pipeline and fitting the resulting
points with a spline curve. We then used this curve for our 
flux calibration. The resulting spectra exhibited a much better 
geocoronal Ly${\alpha}$ subtraction. Although in some cases the signal 
to noise was improved significantly, the overall signal to noise in
the mean spectrum from both visits was less (S/N${\simeq}25$) than the 
one produced from the STScI pipeline (S/N${\simeq}35$). 
Table~\ref{tab:observations} lists all the individual exposures
obtained and the corresponding orbital phase at the middle of each 
exposure.

\begin{table}
\begin{center}
\caption{Log of STIS exposures.}
\begin{center}
\begin{tabular}{cccc}

\hline
Spectrum & \multicolumn{2}{c}{Mid-exposure} & Exposure \\
         & Julian Date & Phase        &    time (s)   \\
\hline
Visit 1 &&&\\
11a & 2451002.1866 & 0.69089 & 600 \\
12a & 2451002.1951 & 0.70286 & 600 \\
12b & 2451002.2023 & 0.71296 & 600 \\
13a & 2451002.2507 & 0.78079 & 600 \\
13b & 2451002.2579 & 0.79089 & 600 \\
13c & 2451002.2651 & 0.80098 & 600 \\
13d & 2451002.2758 & 0.81603 & 600 \\
14a & 2451002.3179 & 0.87504 & 600 \\
14b & 2451002.3251 & 0.88513 & 600 \\
14c & 2451002.3323 & 0.89523 & 600 \\
14d & 2451002.3395 & 0.90533 & 600 \\
15a & 2451002.3856 & 0.96996 & 600 \\
15b & 2451002.3928 & 0.98006 & 600 \\
15c & 2451002.3000 & 0.99016 & 600 \\
15d & 2451002.4072 & 1.00025 & 600 \\
Visit 2 &&&\\
21a & 2451013.0168 & 0.88017 & 600 \\
22a & 2451013.0254 & 0.89213 & 600 \\
22b & 2451013.0326 & 0.90222 & 600 \\
23a & 2451013.0725 & 0.95824 & 600 \\
23b & 2451013.0797 & 0.96834 & 600 \\
23c & 2451013.0869 & 0.97843 & 600 \\
23d & 2451013.0976 & 0.99348 & 600 \\
24a & 2451013.1397 & 1.05257 & 600 \\
24b & 2451013.1469 & 1.06263 & 600 \\
24c & 2451013.1541 & 1.07273 & 600 \\
24d & 2451013.1613 & 1.08283 & 600 \\
25a & 2451013.2075 & 1.14751 & 600 \\
25b & 2451013.2147 & 1.15761 & 600 \\
25c & 2451013.2219 & 1.16770 & 600 \\
25d & 2451013.2291 & 1.17780 & 600 \\

\hline
\end{tabular}
\end{center}
\protect\label{tab:observations}
\end{center}
\end{table}

%------------------------------------------------------------------------------------------

\section{The mean spectrum}
\label{sec:mean_spec}
\begin{figure*}
\begin{minipage}{170mm}
\vspace*{10cm}          
\includegraphics{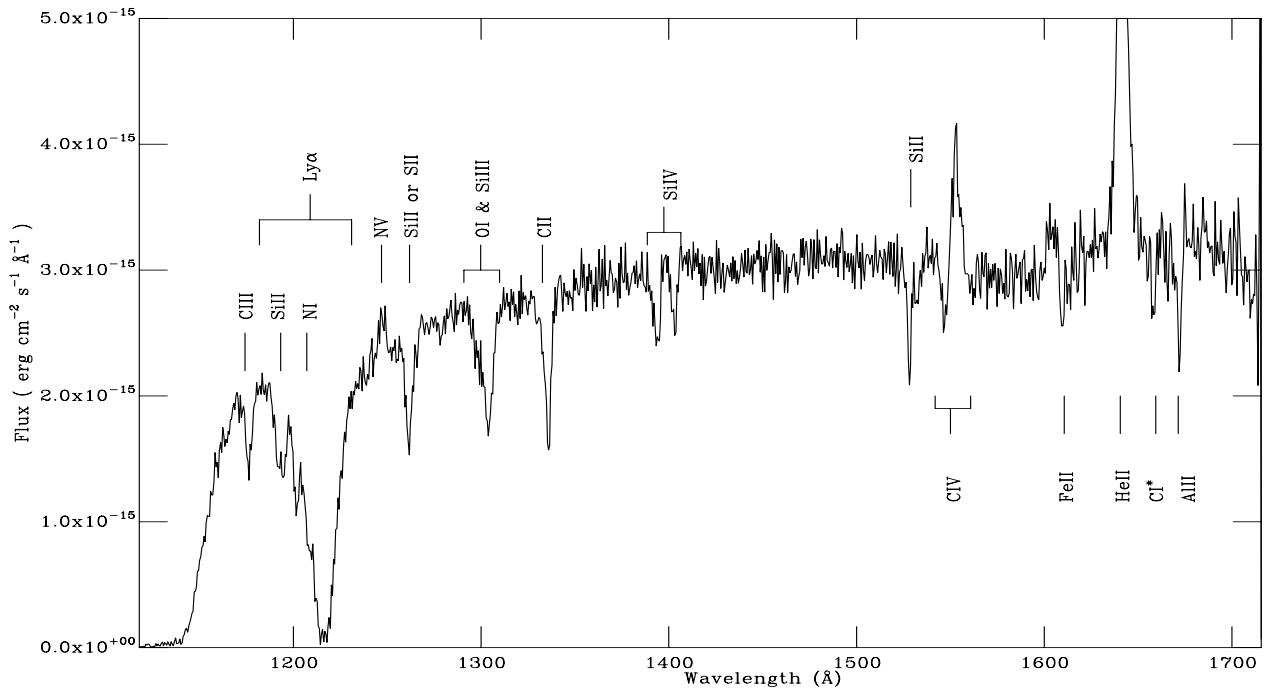}
\caption{The mean spectrum of AC211 produced from observations taken
during both visits.}
\label{fig:spectrum}
\end{minipage}
\end{figure*}

Our observations covered from phase
0.69 up to the main eclipse at phase 0.0 for visit 1, and from 
phase 0.88 through eclipse to phase 0.18 for visit 2.
The combined mean spectrum produced from both visits can be seen in
Fig~\ref{fig:spectrum}. The main features are strong 
{\heii} (1640{\AA}) emission, weak N{\sc v} (1247{\AA}) emission, 
a broad Ly${\alpha}$ absorption region around 1180-1240{\AA} and a weak 
{\civ} (1550{\AA}) P Cygni line. 
There are also three absorption features superposed on the 
Ly${\alpha}$ feature. These are C{\sc iii} (1176{\AA}), Si{\sc ii} 
(1192{\AA}) and N{\sc i} (1202{\AA}). 
There is a line at 1260{\AA} which could be either Si{\sc ii} or
S{\sc ii} and a feature at 1300{\AA} which could be due to either 
Si{\sc iii} or O{\sc i}. C{\sc ii} (1335{\AA}) absorption is present
well as the Si{\sc iv} (1400{\AA}) absorption doublet. 
In addition 
several narrow lines are also present in the spectrum like Si{\sc ii} 
(1528{\AA}), Fe{\sc ii} (1610{\AA}), C{\sc i}$^{\star}$ (1658{\AA})
and Al{\sc ii} (1672{\AA}), which we believe are interstellar in origin
(see Sect. \ref{sec:inter}). 
Several of these spectral features including 
Si{\sc iv} (1400{\AA}) and {\civ} (1550{\AA}) are seen for the first
time in this system. Table~\ref{tab:lineids} lists all the identified 
lines along with measurements of their equivalent widths and fluxes.

%------------------------------------------------------------------------------------------

\section{The UV continuum and eclipse}
\label{sec:continuum}

The average value of our continuum flux measurements is about 
$3{\times}10^{-15}$erg~cm$^{-2}$s$^{-1}${\AA}$^{-1}$, but modulations
by a factor of 2 are observed above and below this value.
Even our brightest measurement is below that observed
by \cite{Downes96} with the {\it HST FOS} 
($8{\times}10^{-15}$erg~cm$^{-2}$s$^{-1}${\AA}$^{-1}$), which could
be indicative of long term variability. 
However, their observations were taken at phases around 
0.4 and investigation of the X-ray light curves presented in \cite{Ioannou01}
(hereafter Paper I) shows that phases around 0.4 also show the highest X-ray flux. 

\begin{figure}
\vspace{7.5cm}          
\includegraphics{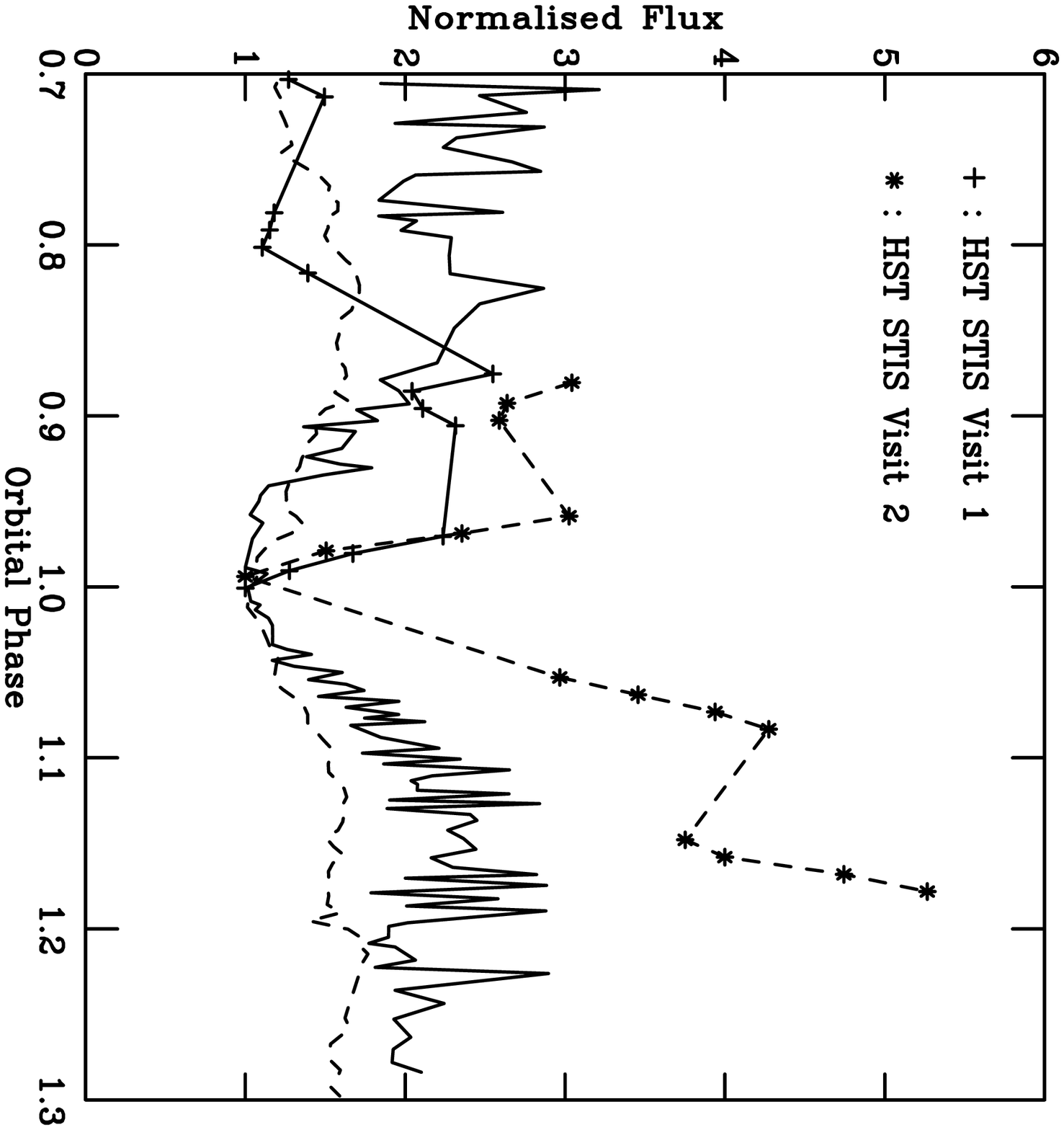}
\caption{The behaviour of the continuum flux 
(in the band of 1420\AA-1510\AA) with orbital phase. 
Overplotted are the 
optical eclipse (solid line) from \cite{Ilovaisky93} and the X-ray
eclipse modeled in Paper I (``96a'', dashed line; note that the flux
includes M15 X-2). 
The optical eclipse was folded on
the ephemeris of \cite{Ilovaisky93}, while the X-ray and UV eclipses
were folded on the ephemeris discussed in Sect. \ref{sec:period}.}
\label{fig:contflx}
\end{figure}

Fig.~\ref{fig:contflx} shows the modulation of continuum flux with
orbital phase as observed from both visits. The eclipse close to phase
1.0 is evident. We attribute the drop in continuum flux around phases 
0.7-0.85 to vertical structure in the accretion disc as discussed
in Paper I. The data were folded on the ephemeris which is discussed
in Sect. \ref{sec:period}.

Despite the fact that our spectra are of poor orbital coverage during 
and around eclipse, we have attempted to compare the radial extent of
the UV emitting region with that of the X-ray and optical emitting regions.
In Paper I we derived a radius for the X-ray emitting region of 
$R{\simeq}0.8R_{L1}$, where $R_{L1}$ is the distance 
from the centre of the disc to the inner Lagrangian point. 
That modelling assumed all of the X-ray flux originated from AC211, but 
varying the amount of contaminating light from M15 X-2 does not change the
derived radius.
This is because its is the width (not the depth) of the eclipse which
constrains the size of the eclipsed object.
Using the same system parameters
for the inclination and mass ratio of the system as in Paper I we find
that the extent of the UV emitting region on the orbital plane of the binary 
must be about $0.3-0.4R_{L1}$.
The reason for this difference in the derived size is driven by the eclipse
widths, which has a FWHM of 0.15 in phase in the X-ray data (Paper I) compared with about
half this value in the UV (see Fig.~\ref{fig:contflx}).

Fig.~\ref{fig:contflx} also allows a direct comparison between the UV and 
optical eclipses, the latter taken from \cite{Ilovaisky93}.
It is evident that the optical eclipse is wider (its FWHM is 0.15 in phase
measured from the data of \citeauthor{Ilovaisky93}). 
The eclipse can only be wider if the optically eclipsed material fills more 
of the Roche-lobe in the plane of the orbit than the material eclipsed in the
UV.
Thus we find that the UV emitting region is smaller, and more centrally 
concentrated than both the X-ray and optical regions.
We must, though, caution that AC211 is a highly variable source, and thus far
we have only obtained observations of part of 2 UV eclipses.

\begin{table}
%\begin{minipage}{170mm}
\begin{center}

\caption{Identifications, equivalent widths and fluxes of UV lines 
in the spectrum of AC211.}
\begin{center}
\begin{tabular}{lccc}

\hline
Element                 & ${\lambda}$ & $W_{\lambda}$$^{a}$ & Flux $({\times}10^{-15})$\\
                        & (\AA)       & (\AA)         & (erg~cm$^{-2}$~s$^{-1}$) \\
\hline
C{\sc iii}$^{b}$              & 1176.3 & $-1.19{\pm}0.13$ &$2.46{\pm}0.27$\\
Si{\sc ii}$^{b,c}$            & 1192.0 & $-1.60{\pm}0.14$ &$3.08{\pm}0.26$\\
                              & 1194.9 &                  &\\
N{\sc i}$^{b}$                & 1201.7 & $-0.75{\pm}0.11$ &$1.17{\pm}0.17$\\
N{\sc v}$^{b}$                & 1247.6 & $ 1.05{\pm}0.09$ &$2.34{\pm}0.20$\\
Si{\sc ii} or S{\sc ii}$^{b}$ & 1261.6 & $-1.56{\pm}0.15$ &$3.88{\pm}0.23$\\
O{\sc i} \& Si{\sc ii}$^{d}$  & 1303.7 & $-2.51{\pm}0.15$ &$7.47{\pm}0.62$\\
C{\sc ii}$^{d}$               & 1336.1 & $-1.91{\pm}0.14$ &$5.78{\pm}0.38$\\
Si{\sc iv}$^{d}$              & 1393.7 & $-0.85{\pm}0.07$ &$2.82{\pm}0.25$\\
                              & 1403.6 & $-0.66{\pm}0.08$ &$2.16{\pm}0.24$\\
Si{\sc ii}$^{d}$              & 1528.2 & $-0.89{\pm}0.06$ &$3.00{\pm}0.29$\\
C{\sc iv}                     & 1550.0 & see table~\ref{tab:civline} &\\
Fe{\sc ii}$^{b}$              & 1609.6 & $-0.37{\pm}0.10$ &$1.13{\pm}0.29$\\
He{\sc ii}$^{d}$              & 1641.6 & $10.95{\pm}1.08$ &$33.5{\pm}1.7$ \\
C{\sc i}$^{b}$                & 1658.2 & $-0.44{\pm}0.16$ &$1.41{\pm}0.37$\\
Al{\sc ii}$^{b}$              & 1672.2 & $-0.74{\pm}0.18$ &$2.32{\pm}0.55$\\

\hline
\end{tabular}
\end{center}
\protect\label{tab:lineids}
\end{center}
%\end{minipage}
$^{a}$ Negative values indicate absorption lines.\\
$^{b}$ The equivalent width and flux values are from
the mean weighted spectrum of both visits.\\
$^{c}$ Total equivalent width and the flux value for both lines of the
Si{\sc ii} doublet.\\
$^{d}$ The equivalent width and flux values are the averages
from individual spectrum measurements. 
\end{table}

%------------------------------------------------------------------------------------------

\section{The {\heii} 1640{\AA} line}
\label{sec:heii}

\begin{figure}
\vspace{7.5cm}    
\includegraphics{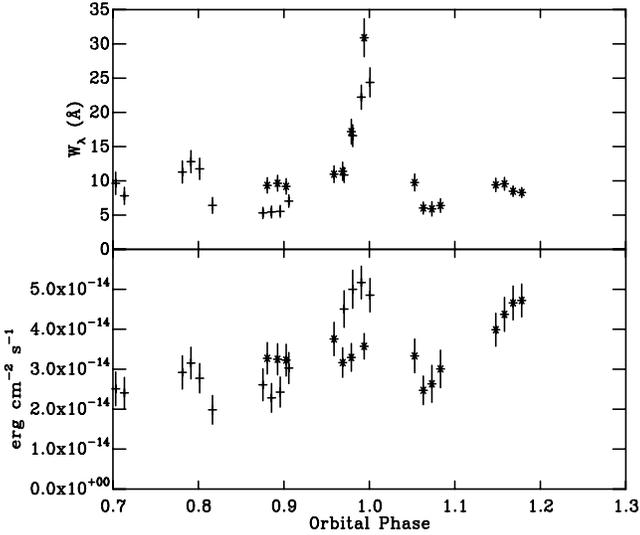}
\caption{The behaviour of the {\heii} 1640{\AA} line. The top panel
shows the change in the equivalent width with orbital phase and the
lower panel the change in flux. Observation made in Visit 1 are
depicted by crosses and those from Visit 2 by stars.}
\label{fig:HeIIflxeqw}
\end{figure}

\begin{figure*}
\begin{minipage}{170mm}          
\includegraphics{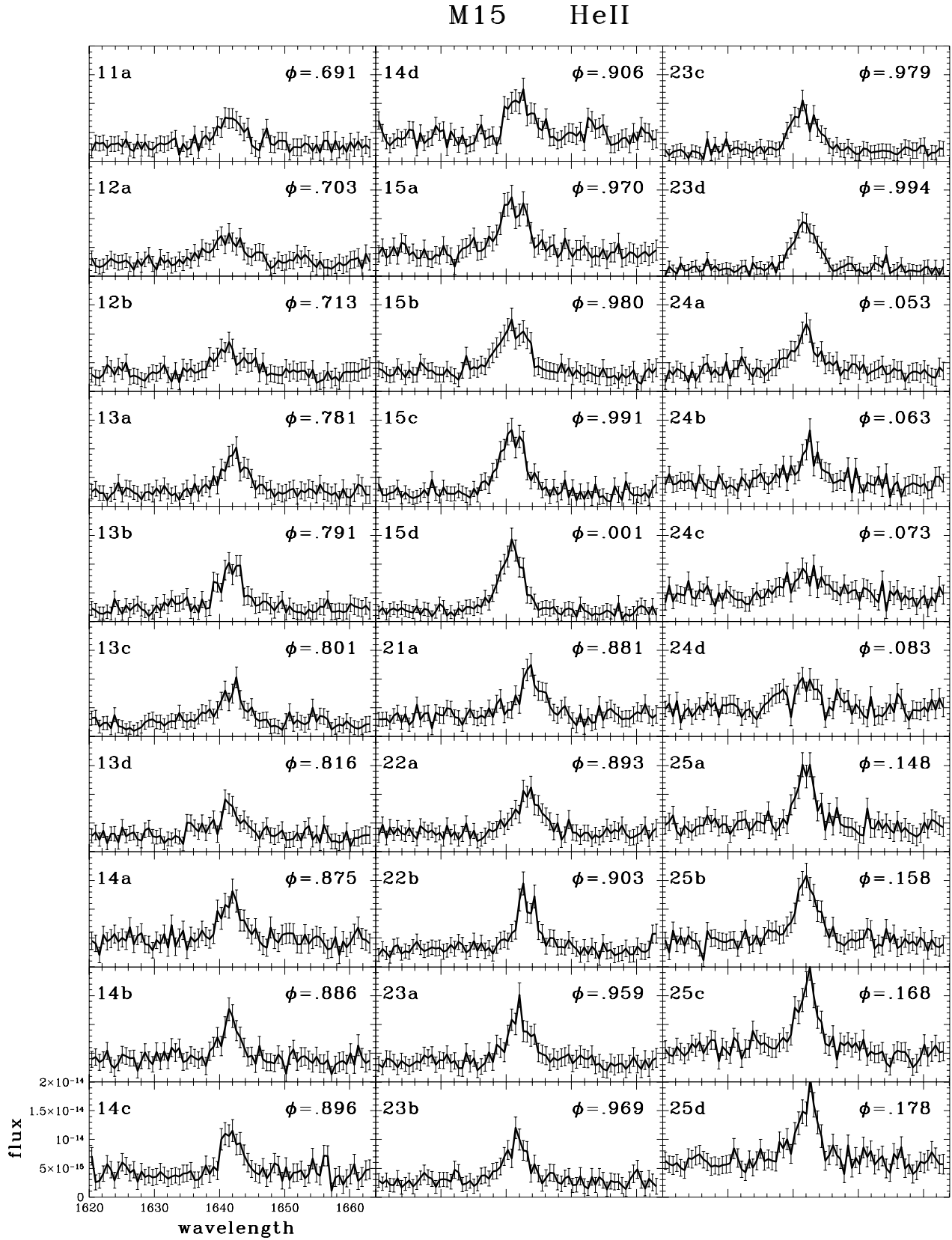}
\vspace{225mm}
\caption{The behaviour of the He{\sc ii} line with orbital
phase. Plots 11a to 15d correspond to visit 1 observations and 21a 
to 25d to visit 2 observations. Wavelength is in {\AA} and flux in 
erg~cm$^{-2}$s$^{-1}$}
\label{fig:HeIIall}
\end{minipage}
\end{figure*}

We have measured the {\heii} flux and equivalent width and plot them 
as functions of orbital phase in in Fig.~\ref{fig:HeIIflxeqw}.
The overall flux level is around a factor 3 lower than in previous 
observations \citep{Naylor88, Naylor92, Downes96} which were all close to 
$1{\times}10^{-13}$erg~cm$^{-2}$s$^{-1}$. 
The flux shows variations at the 50\% level as a function of time, but these
variations do not seem to correlate with orbital phase or between the
same phases in the two different visits. 
Furthermore we do not observe 
any reduction of the {\heii} flux during the eclipse. 
The equivalent width of the {\heii} line rises sharply during the
eclipse, but this is explained by a decrease in continuum
flux (Fig.~\ref{fig:contflx}).
All of the {\heii} line 
profiles from all of our spectra can be seen in Fig.~\ref{fig:HeIIall}. 
It is evident that the {\heii} profile is complex and changes 
continuously. Most importantly it does not show consistent variations 
with orbital phase and different profiles are observed even when spectra
from the two visits correspond to the same orbital phase.

The absence of an eclipse or a strong orbital modulation of {\heii}
flux is an indication that the {\heii} emission originates in a region 
of the system that is not affected by the secondary star or the 
accretion disc. A possible scenario is that it originates in material 
with a vertical extent that is larger 
than the secondary star and which is X-ray heated by the central source and 
the inner accretion disc.
This is in agreement with the geometrical interpretation of
\cite{Naylor92} who suggested that the {\heii} recombination line is
produced in the fully ionized coronal gas of the ADC.

\section{The {\civ} 1550{\AA} line: Estimating the mass loss}
\label{sec:loss}

\begin{figure}
\vspace{8cm}    
\includegraphics{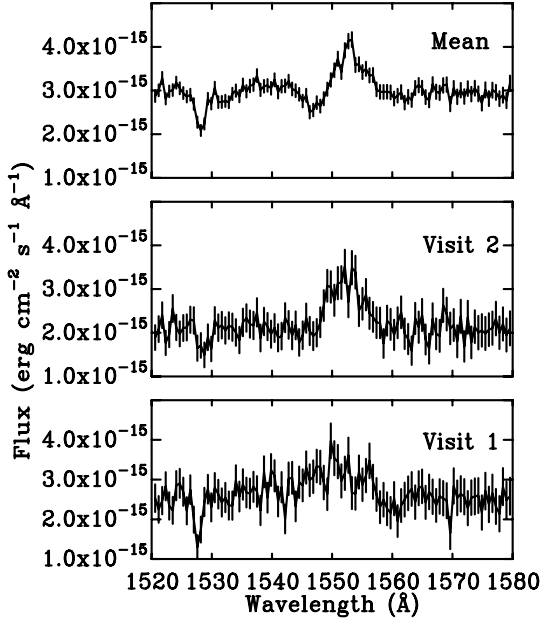}
\label{fig:CIVmean}
\caption{The above figure shows the absence of the absorption
component of the 1550\AA {\civ} line during eclipse. The spectra in
the lower two panels are the weighted mean averaged spectra created
from the observations closest or during eclipse. The top panel
shows the weighted mean spectrum of all our observations with the
absorption component clearly visible.}
\end{figure}

The {\civ} line exhibits a P Cygni profile with a
weak blue-shifted absorption component and a stronger
emission component. A P Cygni profile indicates the existence of a
wind in the system, where the blue shifted absorption profile arises 
from the presence of cool material that is situated in our line of 
sight in front of the hot regions of the accretion disc and traveling
towards us. 
Further evidence that this line comes from an extended wind region
is provided by the behaviour through eclipse (Fig.~\ref{fig:CIVmean}).
Despite the disappearance of the {\civ} absorption component we still
see relatively strong {\civ} emission. This is indicative of the
{\civ} originating in an extended region which is not occulted by the
secondary star.

By measuring the extent of the blue-shifted component we 
can measure the wind velocity in our line of sight.
It is about 1350 km s$^{-1}$ 
for Visit 1 and about 1750 km s$^{-1}$ for Visit 2. We also observe
that in Visit 1 the emission component is sometimes absent in our
individual spectra, while in Visit 2 the inverse situation exists with
the emission component being stronger and the absorption component
being absent in some spectra. Table~\ref{tab:civline} lists the
equivalent widths fluxes and wind velocities as measured from the
weighted mean spectra of each visit.

It should also be noted that \cite{Ilovaisky89} also found
evidence of a  wind by looking at H${\beta}$  absorption profiles. The
wind velocities that  he measured were of the  order of ${\simeq}800$ km
s$^{-1}$.   In comparison,  the wind  velocities found  in cataclysmic
variables are much faster than the wind velocities we find in AC211,
with  the width  of  the  absorption component  in  CVs stretching  to
3000-5000 km s$^{-1}$ \citep{Warner95}.
The velocities are similar to those in Wolf-Rayet (WR) stars 
\citep{vanderHucht01}, including the WR X-ray binary Cyg X-3 
\citep{vanKerkwijk96}, but the mass loss rate we derive below for
AC211 is orders of magnitude lower than in WR stars.

We have used the method employed by \cite{Hassall83} to estimate
the mass loss rate, $\dot{M}_{L}$. 
The mass loss rate inferred from a {\civ} line showing a P Cygni profile 
for a wind with spherical geometry can be
determined according to the equation:

\begin{equation}
\dot{M}_{L}=1.1{\times}10^{-18}\frac{{\tau}Ru^{2}C}{f_{i}A{\lambda}_{0}f} 
\rm{M_{\odot}yr^{-1}}
\label{eq:massloss}
\end{equation}

Here ${\tau}$ is the optical depth and is determined from the blue
shifted absorption component, $R$ is the effective radius in units of 
solar radii, $u$ is the wind velocity in km s$^{-1}$ as measured from the
extent of the absorption profile. $A$ is the abundance of the line
element, ${\lambda}_{0}$=1550{\AA} is the rest wavelength of the line
and $f$ is the line oscillator strength. The value for the constant
$C$ is determined from the blue shifted absorption profile
\citep{Conti80} and like \cite{Hassall83} we have chosen a value of 
$C$=0.3. $f_{i}$ is the fraction of the carbon element in the form 
of {\civ} in the wind. Assuming that all of Carbon in the wind is 
present as {\civ} we have used a value of $f_{i}$=1. We have set the 
effective radius equal to the accretion disc radius $(R=1.7{\rm R_{\odot}})$ 
determined in Paper I. The value for the oscillator
$f$=0.2303 was taken from \cite{Morton91} and we also assume a solar
abundance of carbon and use a value of $A$=0.0004 \citep{Morton91}.
Using the above values we find a mass loss rate of
$7{\times}10^{-12}${\Mdot} for Visit 1 and 
$1{\times}10^{-11}${\Mdot} for the higher wind velocity
corresponding to Visit 2. 

These mass loss rates are a factor of $10^4$ smaller than those suggested
by some models for AC211.
\cite{Homer98} found evidence of a period derivative in the orbital
period of AC211 of the order of $9{\times}10^{-7}$yr$^{-1}$. To explain
this change in period they concluded that the mass loss rate from the
system must be of the order of a few ${\times}10^{-7}${\Mdot}.
\cite{Bailyn89} also suggested a similar amount of material was being
thrown out of the system from the $L_{2}$ point in order to explain
the large blue shift of ${\simeq}100$ km s$^{-1}$ of the He{\sc i} line
as originally reported by \cite{Naylor88}. 

In an attempt to reconcile these mass-loss rates we examined the 
assumptions we had made.
The use of the {\civ} profile to determine
the mass loss rate in AC211 requires that we know the value for the
abundance of carbon in the system. The metallicity of M15
has a value of $[Fe/H]=-2.17$ \citep{Djorgovski93b}. However, even by
lowering the carbon abundance by two orders of magnitude we still do
not get the very high mass loss rates of ${\sim}{\times}10^{-7}${\Mdot}.
In addition we know that the value of $f_{i}=1$ is
only an approximation, since a C{\sc iii} (1176{\AA}) line is present at 
certain phases in our individual spectra. However, it is evident even 
by this crude determination of $\dot{M}_{L}$, that the wind giving rise to 
this {\civ} profile cannot produce the mass loss required 
by \cite{Homer98} and \cite{Bailyn89}.
Also, it should be noted that our above estimates where made with
the assumption that the wind has a spherical geometry. If the wind in
AC211 is bipolar then higher mass transfer rates would be possible.

\begin{table}
\begin{center}
\caption{Equivalent width, flux and measured wind velocities for
each visit from the {\civ} 1550{\AA} line. The measurements were made
using the weighted mean spectra of each visit.}
\begin{center}
\begin{tabular}{lcccc}

\hline
Line      & Visit & $W_{\lambda}$ & Flux $({\times}10^{-15})$ & Wind \\
Component &  No   &     \AA       &  erg~cm$^{-2}$~s$^{-1}$   & km s$^{-1}$ \\
\hline
Absorption & Visit 1 & $0.51{\pm}0.12$ & $1.43{\pm}0.34$& 1350\\
           & Visit 2 & $1.43{\pm}0.34$ & $1.36{\pm}0.34$& 1750\\   
\\
Emission   & Visit 1 & $1.58{\pm}0.18$ & $4.42{\pm}0.49$& \\
           & Visit 2 & $1.72{\pm}0.17$ & $5.46{\pm}0.54$& \\   

\hline
\end{tabular}
\end{center}
\protect\label{tab:civline}
\end{center} 
\end{table}

%------------------------------------------------------------------------------------------

\section{Is there a period change in AC211?}
\label{sec:period}

\begin{figure}
\vspace{7cm}    
\includegraphics{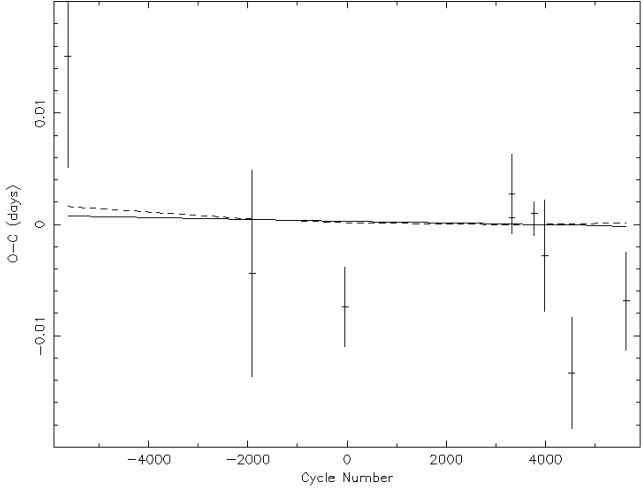}
\label{fig:o-c}
\caption{The O-C diagram for AC211. The solid line shows the best
linear fit and the dashed line shows the best quadratic fit.}
\end{figure}

\begin{table}
\begin{center}

\caption{X-ray observations used to calculate the ephemeris.}
\begin{center}
\begin{tabular}{lll}

\hline

  Mid-Eclipse   &  Error   & Satellite/ \\
(JD +2450000.0) &  (days)  & Instrument \\
\hline

 43469.375877   & 0.009982 &  EXOSAT    \\   
 46120.361216   & 0.009269 &  EXOSAT    \\
 47457.268883   & 0.003565 &  Ginga     \\
 49855.162059   & 0.003565 &  ASCA/GIS  \\
 49855.159920   & 0.001426 &  ASCA/SIS  \\ 
 50178.158      & 0.001    &  RXTE/PCA  \\
 50178.870      & 0.001    &  RXTE/PCA  \\
 50329.314230   & 0.004991 &  RXTE/ASM  \\
 50714.333947   & 0.004991 &  RXTE/ASM  \\
 51498.6614     & 0.0044   &  BeppoSAX  \\
 
\hline
\end{tabular}
\end{center}
\protect\label{tab:xobs}
\end{center}
\end{table}

The low mass-loss rate estimates calculated from the {\civ} P-Cygni
profile are in direct contrast with the large mass-loss rate calculated by
\cite{Homer98}. Their extremely high mass-loss
rate was based on their conclusion that AC211 has a large period
derivative. This prompted us to re-investigate the ephemeris question.
We combined the original eclipse timings of
\cite{Homer98} with additional X-ray observations (Paper I) and 
recalculated the ephemeris of AC211. We based our ephemeris
calculation solely on the X-ray eclipse timings and we removed the
optical CCD points that were originally used by Homer \&
Charles. However, we emphasize that it is the addition of the new X-ray
points and not the exclusion of the CCD optical points, which is
responsible for the difference between the two ephemerides.
Table~\ref{tab:xobs} lists all the points used in our ephemeris calculation.
We found that the data could be folded on the following linear
ephemeris, with the uncertainty in the last figure shown in brackets.
\begin{equation}
T_{0}= HJD~2447484.371(5) + 0.71301(2){\times}E
\end{equation}
The above ephemeris is in very good agreement with the ephemeris of
\cite{Ilovaisky93}.
Our ephemeris residuals (O-C values) are shown in Fig.6.
We also attempted a quadratic fit to the data. The best quadratic fit
found corresponds to the following ephemeris (in days).
\begin{equation}
O-C = 1.4{\times}10^{-4} - 1.3{\times}10^{-7}E + 2.3{\times}10^{-11}{\times}E^{2}
\end{equation}
However, the above quadratic term is significant only at the 7\% level!
Thus our expanded data set provides
no statistically-significant evidence for a period change in AC211.

%------------------------------------------------------------------------------------------

\section{Which lines are interstellar?}
\label{sec:inter}

Apart from AC211 the STIS MAMA detector recorded the spectra of 
a number of other field stars. Comparing AC211 with the
brightest other star on each visit, we can make a
selection of which lines are due to the interstellar medium and which
are intrinsic to AC211. We find that apart from the {\heii}, {\civ},
Si{\sc iv} and N{\sc v} lines all the other lines which are present in
AC211 are also present in the spectra of the two other field stars. 
The profiles of Si{\sc ii}, Fe{\sc ii} and Al{\sc ii} are narrow profile, 
which is indicative of interstellar lines. 
However, Ly${\alpha}$, 
as well as the absorption features at 1261{\AA} (Si{\sc ii} or 
S{\sc ii}), 1303{\AA} (O{\sc i} and Si{\sc iii} blend) and 1336{\AA} 
(C{\sc ii}) have a broader profile in AC211 than in 
the spectra of the background field stars. This indicates that some 
of the absorption seen in these lines is intrinsic to AC211; indeed
the  Ly${\alpha}$ profile varies with phase.
 
Since the Si{\sc ii} absorption line at 1528{\AA} shows none of
these problems, we can use it to estimate the column of Si towards AC211. 
\cite{Savage91} show that the apparent column density 
per unit velocity is given by:

\begin{equation}
N_{a}=\frac{3.768{\times}10^{14}}{f{\lambda}}{\int}ln\left[\frac{I_{0}(u)}{I(u)}\right]du
\label{eq:absorption}
\end{equation}

where $f$ is the absorption oscillator strength of the line,
${\lambda}$ is the wavelength in \AA\ and $I_{0}(u), I(u)$ are
the unabsorbed and absorbed line intensities respectively. $N$ is
given here in units of atoms cm$^{-2}$~km s$^{-1}$. \cite{Savage91}
also show that apparent column densities calculated in this fashion are
instrumentally blurred but non-the-less accurate as long as the line
is resolved or unless the integration in Eq.~\ref{eq:absorption} is 
performed over an entire line which has no saturated components.
Normalizing our absorption profile to the continuum level and using an
oscillator strength value for Si{\sc ii} of $f=0.2303$ \citep{Morton91}, we 
calculate a value for $N_{Si}=(1.4{\pm}0.1){\times}10^{15}$cm$^{-2}$
towards AC211. 

\cite{Shull85} calculated galactic interstellar abundances using 
{\it IUE} for 244 early type stars for
which a number of them (68) are halo objects. Of these, 25 lie in the
direction of M15 ($l$=65, $b$=-27). Plotting the $log(N_{Si})$
against the $log(N_{H})$ values of the 25 stars and by fitting a
straight line through the points we determine a value of 
$N_{H}=(3.5{\pm}0.2){\times}10^{20}$cm$^{-2}$. This is in excellent 
agreement with the $N_{H}$ value calculated in Paper I using an 
interstellar reddening value of $E_{(B-V)}=0.05$ from 
\cite{Djorgovski93a}.

Using the same method, we have also measured the silicon and hydrogen
column densities by for two other stars,
whose spectra were also recorded on the MAMA CCD detector (the
brightest, aside from AC211, on each visit).
All the equivalent widths and derived column densities are given 
in Table~\ref{tab:columns}.

\begin{table}
\begin{center}
\caption{Column densities for Silicon and Hydrogen as derived by using
the Si{\sc ii} 1528{\AA} line.}
\begin{center}
\begin{tabular}{llll}

\hline
Star   & $W_{\lambda}$   & $N_{Si}({\times}10^{15})$ &$N_{H}({\times}10^{20})$\\
       & (\AA)           &       cm$^{-2}$          &       cm$^{-2}$       \\
\hline
AC211  & $0.89{\pm}0.06$     & $1.4{\pm}0.1$            & $3.5{\pm}0.2$ \\
Star 1 & $0.89{\pm}0.14^{a}$ & $1.6{\pm}0.1$            & $4.1{\pm}0.3$ \\
Star 2 & $0.88{\pm}0.22^{a}$ & $0.89{\pm}0.14$          & $1.8{\pm}0.4$ \\   

\hline
\end{tabular}
\end{center}
\protect\label{tab:columns}
\end{center}
$^{a}$ The equivalent width is measured from the mean weighted
spectrum of the star.\\ 
\end{table}

%------------------------------------------------------------------------------------------

\section{Discussion}

The HST STIS observations show that
AC211 has a strong UV continuum with a host of absorption lines, many
of which are interstellar in origin (Sect. \ref{sec:mean_spec}
and \ref{sec:inter}).
In addition there is a highly variable emission from {\heii} and a {\civ}
P-Cygni profile (Sect. \ref{sec:heii} and \ref{sec:loss}).
Since the {\heii} line is uneclipsed, we conclude it must come from 
a relatively large volume above and below the disc (Sect. \ref{sec:heii}).
From the {\civ} line we can estimate the mass-loss from the system as 
$\sim10^{-11}${\Mdot} (Sect. \ref{sec:loss}) much lower than that 
suggested by the orbital period derivative of \cite{Homer98}.
However, the inclusion of new data implies the period derivative 
is no longer significant, and thus that the derived mass-loss rate is probably
correct (Sect. \ref{sec:period}).
These facts all fit nicely within the normal expectation for an X-ray
binary.
To progress further, we must accept that AC211 is a high inclination
eclipsing system, in which the X-ray source is hidden behind the rim of the
accretion disc, and that the X-rays originate from an accretion disc corona.
This has long been the standard model for AC211, albeit with problems.
The discovery of M15 X-2 removes the major objection to this model, which
is how such a system could show an X-ray burst (see, for example, the
discussion in Paper 1).
The simplest assumption now is that M15 X-2 was responsible for the burst,
and that the compact object in AC211 is indeed hidden.

At first sight the eclipses seem to fit easily into this model.
The optical eclipse is wide, as one would expect from the eclipse of an
accretion disc, since the entire disc will be X-ray heated to temperatures
which emit significant radiation in the optical.
The UV eclipse is narrower, again expected as only the inner regions of the
accretion disc emit significant UV light.
Finally, the X-ray eclipse is wide, since it is again of an extended object,
the corona.
We note in passing that the flux at the bottom of the X-ray eclipse is
consistent with that from X-2, i.e. it could well be that the eclipse of
AC211 is total.
Although this means the accretion disc corona is smaller than previously
thought, it does not mean the X-ray source is point like.
If it were, the eclipse ingress and egress would be vertical, instead their
slopes imply the X-ray emitting region is similar in size to the
secondary star.

The problem with the UV eclipse is twofold.
First, it is not total; one would expect a small, compact UV region to
be totally eclipsed.
This is backed up by the fact that the UV ingress and egress is not vertical,
but our poor phase coverage make this result uncertain.
Second, one might not expect to see the inner disc at all; it should be
hidden behind the disc rim.
Thus it appears that the UV emitting region is significantly extended, but
how, or why remains unclear.
The only way around this conundrum would be to invoke variability, since
the X-ray, optical and UV eclipses are not simultaneous.
Such a solution is is unsatisfactory, though, since the X-ray and
optical eclipses are relatively stable in width (Paper I and 
\citealp{Ilovaisky93}), and the phases in common
between our two HST visits are give similar UV fluxes.

%------------------------------------------------------------------------------------------
\section{Conclusions}

The STIS UV observations presented here have shown that, although the
eclipse in AC211 implies that the UV source is more centrally
concentrated than either the X-ray or the optical source, it
cannot be explained by a simple disc model.

The interstellar UV lines imply an $N_{H}=3.5{\pm}0.2{\times}10^{20}$
cm$^{-2}$, whilst the {\civ} P~Cygni profile implies a mass-loss rate
of \.M$_{L}{\sim}1{\times}10^{-11}${\Mdot}. This is at variance
with the mass-loss rates implied by \cite{Homer98}, 
but the addition of new X-ray data to the ephemeris
calculation removes the need for a large period change, which in
turned implied the high mass-loss rate.

%------------------------------------------------------------------------------------------
\begin{acknowledgements}

We would like to thank Lee  Homer for providing the X-ray data used in
the   ephemeris  calculations.   LvZ   acknowledges  the   support  of
scholarships  from  the  Vatican  Observatory, the  National  Research
Foundation  (South  Africa), the  University  of  Cape  Town, and  the
Overseas  Research Studentship  scheme (UK).  TN was  in receipt  of a
PPARC advanced fellowship  when the majority of this  work was carried
out.   The work  presented  here  is based  on  observations with  the
NASA/ESA  Hubble  Space Telescope,  obtained  at  the Space  Telescope
Science   Institute,  which   is  operated   by  the   Association  of
Universities  for Research  in  Astronomy, Inc.,  under NASA  contract
NAS5-26555.
\end{acknowledgements}

%------------------------------------------------------------------------------------------

\bibliographystyle{aa}

%%%%%%%%%%%%%%%%%%%%%%%%%%%%%%%%%%%%%%%%%%%%%%%%%%%%%%%%%%%%%%%%%%%%%%

\end{document}